\pdfoutput=1   
\documentclass[11pt]{article}

\usepackage[utf8]{inputenc}
\usepackage[T1]{fontenc}
\usepackage{mathptmx}          
\usepackage[scaled=0.92]{helvet}
\usepackage{courier}

\usepackage[letterpaper,left=1.15in,right=1.15in,top=1.1in,bottom=1.15in]{geometry}
\usepackage{amsmath,amssymb}
\usepackage{graphicx}
\usepackage{booktabs}
\usepackage{tabularx}
\usepackage{array}
\usepackage{multirow}
\usepackage{makecell}
\usepackage{ragged2e}
\usepackage{enumitem}
\usepackage{microtype}
\usepackage{authblk}
\usepackage{titlesec}
\usepackage[font=small,labelfont=bf,labelsep=period,justification=justified,singlelinecheck=false]{caption}
\usepackage{float}
\usepackage[dvipsnames]{xcolor}
\usepackage[colorlinks=true,allcolors=black,breaklinks=true]{hyperref}
\titleformat{\section}{\normalfont\large\bfseries}{\thesection}{0.7em}{}
\titleformat{\subsection}{\normalfont\normalsize\bfseries\itshape}{\thesubsection}{0.7em}{}
\titlespacing*{\section}{0pt}{1.6ex plus .6ex minus .2ex}{0.9ex plus .2ex}
\titlespacing*{\subsection}{0pt}{1.3ex plus .5ex minus .2ex}{0.6ex plus .2ex}

\newcommand{\cit}[1]{\textsuperscript{#1}}
\newcommand{\ra}{$\rightarrow$}
\newcommand{\ub}{\_\allowbreak}   

\newlength{\figbleed}
\newcommand{\widefig}[1]{%
  \makebox[\textwidth][c]{%
    \includegraphics[width=\dimexpr\textwidth+2\figbleed\relax]{#1}}}
\newcommand{\iv}{\textit{in vivo}}

\newcolumntype{L}[1]{>{\RaggedRight\arraybackslash}p{#1}}
\newcolumntype{Y}{>{\RaggedRight\arraybackslash}X}
\newcommand{\thd}[1]{\textbf{#1}}

\begin{document}

\title{\bfseries CircuitATLAS: Agentic reasoning over a systems neuroscience
knowledge graph for target discovery in circuitopathies}

\author[ ]{Gabriel Ocana-Santero\textsuperscript{*}}
\author[ ]{Marko Tvrdic}
\affil[ ]{\normalsize Exin Therapeutics, Inc., 2261 Market Street STE 22614,
San Francisco, California 94114, United States}
\affil[ ]{\normalsize \textsuperscript{*}Correspondence: \texttt{gabriel@exintherapeutics.com}}
\date{}

\maketitle
\thispagestyle{empty}

\begin{center}
\begin{minipage}{0.92\textwidth}
{\bfseries\large Abstract}\\[0.4em]
\small
Drug discovery for neurological disease has traditionally centered on the
molecules altered by disease. But the molecular changes that cause pathology are
not necessarily the only or even the most effective points from which to
reverse it. Here, we ask which otherwise unaltered molecular control points can
be engaged to restore pathological neural circuits toward functional states. For
this, we present CircuitATLAS, a provenance-grounded systems-neuroscience
knowledge graph and agentic framework for target discovery in circuitopathies.
CircuitATLAS structures literature-derived relationships across diseases,
phenotypes, electrophysiology, circuits, brain regions, cell types and molecular
effectors, while deliberately excluding direct disease-gene and disease-protein
edges from target generation to reduce shortcut reasoning. The current graph
contains 3.83 million nodes and 7.66 million edges, including 5.31 million
LLM-extracted relations, and can incorporate structured datasets such as the
Human Cell Atlas and newly generated multimodal \iv{} measurements. Then we
introduce a new agentic reasoning workflow that reasons from measurable disease
phenotypes through their circuit and cellular substrates to molecular
interventions, therapeutic feasibility and clinical constraints. Finally, we
introduce a human-governed \iv{} lab-in-the-loop architecture linking hypothesis
generation to experimental iteration. Within this framework an agent nominated
ATP1A3, the neuronal $\alpha$3 Na$^{+}$/K$^{+}$-ATPase, as a control point on
cortical excitability; interneuron-restricted expression of ATP1A3 abolished the
beta- and gamma-band response to a focal 4-aminopyridine challenge \iv{}, and
the validated target was then carried into a structure-guided small-molecule
campaign that terminates in a defined assay to resolve the direction of
modulation.
Together, CircuitATLAS provides a framework for discovering therapeutics based
not only on what is molecularly disrupted in disease, but on what can be
controlled to restore circuit function.
\end{minipage}
\end{center}

\vspace{1.2em}

\section{Introduction}

Neurological and neuropsychiatric disorders are commonly classified by clinical
presentation, molecular pathology, or genetic cause, yet their resulting
phenotypes ultimately emerge from the dynamics of interacting neural
populations\cit{1--3}. Seizures, bradykinesia, cognitive impairment, sensory
hypersensitivity, and impaired social behavior are not properties of isolated
genes or cells. Instead, these phenotypes emerge from aberrant activity within
and across neural circuits composed of distinct cell populations and brain
regions, operating over multiple temporal scales\cit{4}. Different molecular
perturbations can converge on similar circuit-level dysfunction\cit{5}, while
interventions targeting shared circuit nodes can produce therapeutic benefit
despite underlying genetic heterogeneity\cit{6}. We therefore view many
neurological and neuropsychiatric disorders as circuitopathies, in which
clinically-relevant phenotypes arise from maladaptive neural dynamics that can
be measured and, in principle, therapeutically modulated.

Most molecular drug-discovery strategies begin by identifying genes or proteins
that are mutated, differentially expressed, or otherwise altered in
disease\cit{7,8}. This approach is powerful when the causal defect is known.
However, the molecule that initiates pathology is not necessarily the most
effective point of intervention for reversing the downstream circuit dysfunction
it produces\cit{9}. A therapeutically useful target may be unaltered in the
disease yet well positioned to rebalance a pathological circuit. Antiseizure
medicines\cit{10}, deep-brain stimulation\cit{11}, responsive
neurostimulation\cit{12}, interneuron-replacement therapies\cit{13}, and
experimental excitability-modifying therapies\cit{14} all illustrate this
principle: circuit function can be improved by targeting control points that are
not themselves the primary cause of disease. This reframes the target-discovery
problem: rather than asking only which molecules are altered in disease, we ask
which molecular effectors can drive a pathological circuit toward a more
functional state.

Answering this question systematically is difficult because systems-neuroscience
knowledge is distributed across biological scales and fragmented across studies.
A disease-associated electrophysiological phenotype may be described in one
study, localized to a circuit in another, attributed to a particular cell
population elsewhere, and connected to a potentially druggable receptor or
pathway only through separate perturbation experiments. Relevant evidence spans
species, disease models, anatomical nomenclatures, recording modalities,
behavioral assays, and intervention classes. Conventional literature search
retrieves these observations largely as isolated documents, while most
biomedical knowledge graphs are organized around disease-gene, drug-target, and
pathway relationships rather than the cross-scale causal chain linking neural
activity to circuits, cell types, and molecular control\cit{15--18}.

Large language models provide a means of structuring heterogeneous scientific
literature at scale by resolving contextual relationships between biological
entities across diverse experimental settings\cit{19,20}. However, unconstrained
language-model reasoning is difficult to audit, because intermediate inference
steps need not be grounded in any retrievable evidence. Scientific target
discovery therefore requires more than
retrieval-augmented generation\cit{21}. Evidence must be represented in a
structured and traceable form, with provenance, uncertainty, and conflicting
findings preserved, so that each hypothesis can be linked back to the
observations that support it\cit{22}. The same framework can incorporate newly
generated public and experimental data, placing new observations directly in the
context of existing evidence. Once organized in a structured and
provenance-linked form, these data can support agentic reasoning across
biological scales and experimental contexts that are difficult to integrate
manually.

Here, we introduce CircuitATLAS (Circuit Agent for Therapeutic Latent Analysis
and Search), a systems-neuroscience knowledge graph and agentic framework for
circuit-based target discovery. CircuitATLAS connects diseases, phenotypes,
electrophysiological features, circuits, brain regions, cell types, molecular
effectors, interventions, disease models, and clinical evidence, while
preserving supporting text, source provenance, confidence, and verification
status. Direct disease-to-gene and disease-to-protein edges are deliberately
excluded from the discovery graph so that candidate targets must be supported by
evidence that they regulate relevant cells, circuits, or activity phenotypes,
rather than by disease association alone. The graph incorporates both
literature-derived knowledge and newly-acquired data, including public human
atlases and in-house multimodal \iv{} measurements. On top of this framework, we
introduce an evidence-gated workflow that identifies disease-relevant activity
phenotypes, maps them to circuits and cell populations, nominates molecular
effectors capable of modulating them, and evaluates therapeutic and clinical
feasibility. We connected this discovery framework with a
human-governed \iv{} lab-in-the-loop in which experiments are prioritized against
practical constraints, 3Rs principles, and mandatory human approval, with
resulting data returned to the graph to update subsequent reasoning. Thus,
CircuitATLAS provides a framework for discovering therapeutic control points
based on what can restore circuit function, rather than restricting the search
to what is molecularly altered in disease.

\section{Structuring systems neuroscience into a knowledge graph using LLMs}

Systems-neuroscience knowledge is fragmented across studies that interrogate
different levels of the same biological process, from clinical phenotypes and
abnormal neural activity to the circuits, cell populations, and molecular
mechanisms that control them. CircuitATLAS was designed to connect these
observations into an explicit, machine-readable representation spanning disease,
phenotype, circuit activity, anatomy, cell type, and molecular function
(Figure~1A).

We chose the scientific literature as the initial substrate rather than
attempting to reconstruct knowledge directly from raw experimental datasets.
This loses some information, including fine-grained measurements, negative
results, and analyses not reported by the authors. However, scientific papers
also provide a processed representation of the underlying data, in which
observations have been quality-controlled, statistically analysed, interpreted
in the context of prior work, and expressed in biologically meaningful terms.
Abstracts were particularly useful because they distill the main conclusions of
a study and generally contain less exploratory interpretation than discussion
sections. Full text was retained when available to recover experimental context
and supporting evidence. We also included preprints from bioRxiv, medRxiv, and
arXiv. These sources introduce the risk of incorporating findings that have not
yet undergone peer review, but substantially improve the recency of the graph.
Because these findings may not yet have undergone peer review, source type and
provenance were preserved explicitly, allowing downstream workflows to
distinguish between peer-reviewed and preprint evidence.

To define entities represented in the graph, we first created a curated
neuroscience lexicon containing 1,701 terms and used it to screen records from
PubMed, PubMed Central, Europe PMC, bioRxiv, medRxiv, and arXiv, retaining 15.9
million after screening. We
then assembled the graph vocabulary from both established resources and
purpose-built lists. Genes were imported from NCBI\cit{23}, transcripts from
Ensembl\cit{24}, proteins from UniProt\cit{25}, signalling pathways from
Reactome\cit{26}, brain regions and neuronal cell types from the Allen Brain
Atlas\cit{27}, and phenotypes from the Human Phenotype Ontology\cit{28}.
Concepts that remain poorly standardized in existing databases (e.g., including
circuits, circuit motifs, brain states, and cellular- and circuit-level
electrophysiological features) were curated specifically for CircuitATLAS. Every
node was represented by a canonical name, a set of synonyms, a node class, and a
universal identifier assigned within the graph, while external database
identifiers were retained when available. This combination allowed standardized
molecular knowledge and less formalized systems-neuroscience concepts to be
represented in the same vocabulary.

Papers were assigned to eight overlapping subcorpora by deterministically
detecting nodes from the relevant classes, rather than by asking an LLM to
classify entire documents. These subcorpora covered diseases, phenotypes,
behavior and cognition, brain regions, cell types, circuit motifs, and cellular-
and circuit-level electrophysiological features. Candidate edges were then
generated separately for each permitted pair of node classes. Whenever two nodes
were co-mentioned within a 400-character window, the surrounding passage was
retained as candidate evidence. Passages referring to the same candidate node
pair were then aggregated across publications. This procedure yielded 57.7
million candidate co-occurrences while preserving reproducible document
selection and candidate generation.

LLMs were used only after this deterministic narrowing step, as semantic
verifiers rather than unconstrained extractors (Figure~1B). Candidate relations
were evaluated by Nova 2 Lite, which classified each relation as supported,
unsupported, or uncertain and returned a confidence score and supporting
quotation. Haiku 4.5 was used as a second quality pass over edges Nova 2 Lite
had already accepted. A subset of 4,050 edges, 50 per relation type, was tested on
Opus 5; in 17.8\% of those edges Opus 5 judged the relationship not to be
clearly stated in those passages. Given the rapid evolution of frontier models,
in this study we opted to retain that degree of variability and instead used a
second verification placed at inference on the few edges a candidate chain
depends on during graph traversal.

The resulting schema contains 23 node types connected by 105 populated relation
types: 81 extracted from the literature and 24 imported from canonical datasets
(Figure~1C). Literature-derived edges capture contextual relationships that are
rarely available in structured databases, such as a cell type contributing to a
circuit motif or an intervention modulating an electrophysiological phenotype.
Canonical edges provide complementary relationships such as
gene-transcript-protein mappings, pathway membership, protein interactions, and
drug-target associations. Direct disease-to-gene and disease-to-protein edges
were deliberately omitted. This prevents target discovery from collapsing onto
familiar disease-associated molecules\cit{29} and instead requires molecular
candidates to be reached through evidence that they act on relevant cell types,
circuits, or activity phenotypes.

Every literature-derived edge retains the supporting quotation (Table~1), source
document, publication year, canonical node identifiers, model verdict, confidence, and
extraction metadata. CircuitATLAS currently contains 3,831,997 nodes and
7,659,618 edges, comprising 5,306,909 literature-extracted edges and 2,352,709
canonical edges (Figure~1D). Because papers are processed independently and
evidence is stored at the edge level, new literature can be incorporated through
incremental runs and released as versioned graph snapshots. CircuitATLAS
transforms the scientific literature into an auditable, updateable evidence
graph that supports retrieval and reasoning across biological scales.

\begin{figure}[!htbp]
\centering
\widefig{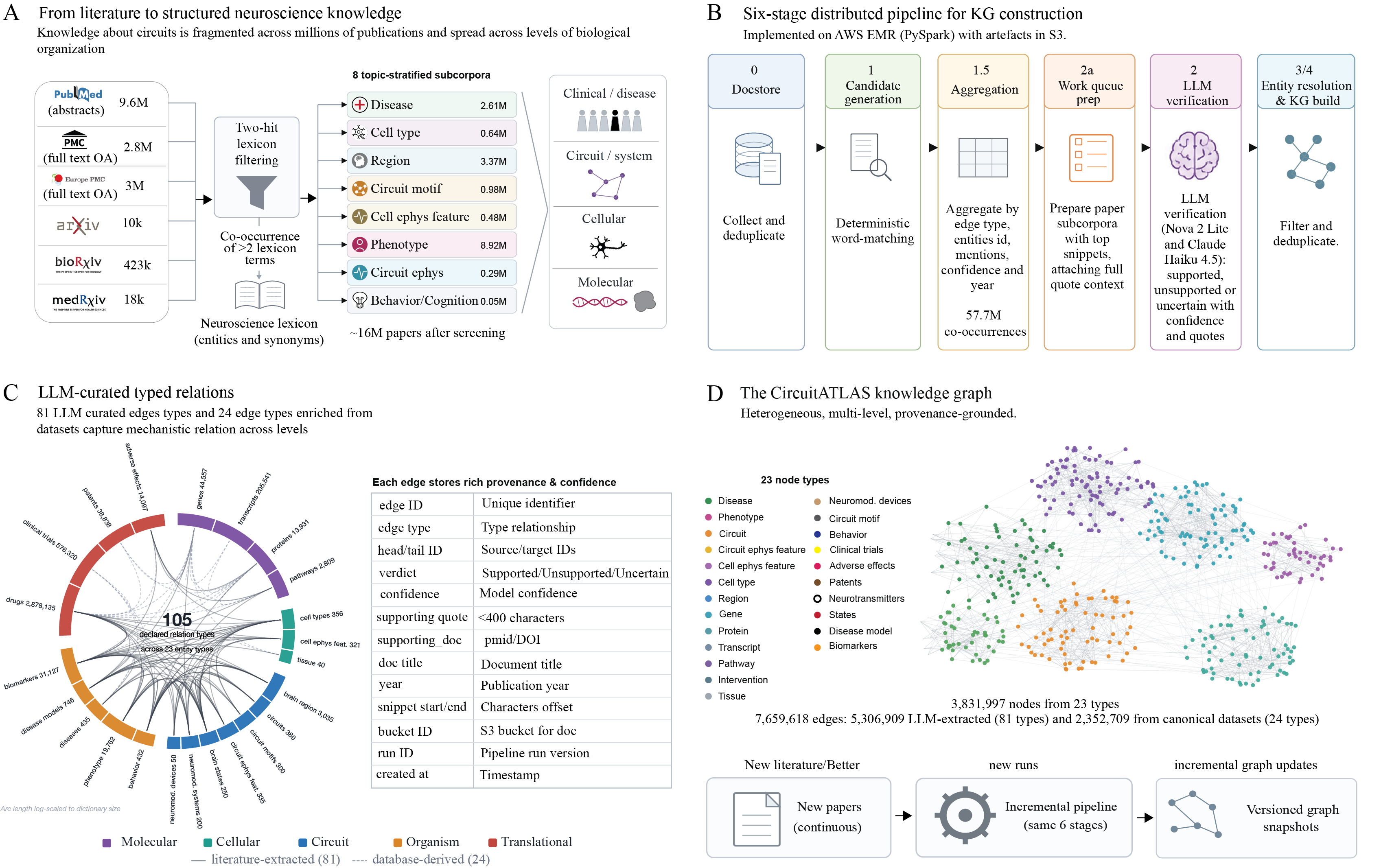}
\caption{\textbf{Construction and structure of the CircuitATLAS knowledge
graph.}
(A) Schematic of the literature-screening workflow. Sources shown are PubMed
abstracts, PubMed Central full text, Europe PMC full text, arXiv, bioRxiv, and
medRxiv. Records are passed through two-hit lexicon filtering and co-occurrence
of at least two lexicon entities, yielding 15.9 million papers after screening.
Retained papers are assigned to eight topic-stratified subcorpora: disease, cell
type, region, circuit motif, cell electrophysiology feature, phenotype, circuit
electrophysiology feature, and behavior/cognition. (B) Schematic of the
six-stage distributed pipeline for knowledge-graph construction. The stages
shown are datastore, candidate generation, aggregation, work-queue preparation,
LLM verification, and entity resolution/KG load. Candidate mentions are
aggregated by edge type, entities, confidence, and year, producing 57.7 million
co-occurrences. The verification stage shows supported, unsupported, or
uncertain classifications with confidence scores and quotations. (C) Circular
summary of LLM-curated typed relations and a table showing example edge fields.
The graph is organized into molecular, cellular, circuit, organism, and
translational levels. The center indicates 105 directed relation types. The
accompanying table lists stored edge attributes, including edge ID, edge type,
head and tail IDs, verdict, confidence, supporting quote, supporting IDs,
document title, year, snippet start/end, bucket ID, run ID, and creation time.
(D) Overview of the CircuitATLAS knowledge graph. Colored node clusters are
shown for 23 node types. The graph contains 3,831,997 nodes and 7,659,618 edges,
including 5,306,909 LLM-extracted edges and 2,352,709 edges from canonical
datasets. A schematic at the bottom shows incorporation of new literature
through incremental pipeline runs and versioned graph snapshots.}
\label{fig:1}
\end{figure}

\begin{table}[!htbp]
\centering
\caption{\textbf{Example LLM-extracted edges}}
\label{tab:1}
\scriptsize
\begin{tabularx}{\textwidth}{L{2.85cm} L{3.15cm} >{\centering\arraybackslash}p{1.05cm} Y >{\RaggedRight\arraybackslash}p{1.25cm}}
\toprule
\thd{Edge type} & \thd{Head \ra{} Tail entity IDs} & \thd{\% support} &
\thd{Evidence quote example} & \thd{PMID}\\
\midrule
\textbf{cell\ub ephys\ub feature \ra{} disease}\newline \textbf{[altered\ub in]} &
Burst afterhyperpolarization/slow AHP (cef276) \ra{} Alzheimer's disease (nd1) &
3/3 &
\textit{``During normal aging and in a mouse model of Alzheimer's disease (AD),
the slow AHP is increased, making neurons less excitable and making learning
more difficult.''} & 15541708\\
\addlinespace[2pt]\midrule
\textbf{drug \ra{} phenotype}\newline \textbf{[ameliorates]} &
Capsaicin (CHEMBL294199) \ra{} Neurodegeneration (HP:0002180) & 17/20 &
\textit{``\ldots capsaicin alleviated other AD-type pathologies, such as tau
hyperphosphorylation, neuroinflammation and neurodegeneration.''} & 32661266\\
\addlinespace[2pt]\midrule
\textbf{biomarker \ra{} disease}\newline \textbf{[associated\ub with]} &
Thyroxine/free T4 (HMDB:0000248) \ra{} Narcolepsy (nd230) & 10/11 &
\textit{``Narcolepsy patients had significantly lower free thyroxine (FT4)
levels in comparison to controls ($p<0.001$).''} & 37898692\\
\addlinespace[2pt]\midrule
\textbf{disease\ub model \ra{} phenotype}\newline \textbf{[recapitulates]} &
Leaner mouse (dm90) \ra{} Dyskinesia (HP:0100660) & 2/2 &
\textit{``The leaner mouse exhibits severe ataxia, paroxysmal dyskinesia, and
absence epilepsy due to a P/Q-type calcium channel mutation.''} & 15514988\\
\bottomrule
\end{tabularx}
\end{table}

\clearpage

\section{CircuitATLAS can integrate public molecular atlases and multimodal
\textit{in vivo} data}

The literature-derived graph captures findings that have already been
interpreted and communicated by the scientific community, but it does not
contain the full resolution of the underlying datasets and lags behind
newly-generated results and datasets. We therefore developed a complementary
route for converting structured public datasets and in-house experimental
analyses directly into the same graph representation. Rather than storing these
datasets as disconnected files, each analysis output is expressed as a typed
edge between canonical CircuitATLAS nodes, together with the quantitative and
experimental context required to interpret it.

We first applied this approach to the adult Human Cell Atlas brain
dataset\cit{30}, which provides single-nucleus transcriptomic profiles across
human brain regions and cell types (Figure~2A). The atlas was used to add two
forms of information that are particularly important for circuit-based target
discovery: (1) the cellular composition of each brain region and (2) the
expression of receptor genes within individual brain regions and cell types. HCA
entities were resolved against the existing CircuitATLAS vocabulary, adding 108
brain-region nodes and 24 cell-type nodes where no equivalent was already
present. The ingestion generated 1,452 HAS\_CELL\_TYPE edges, 99,701
region-to-receptor EXPRESSES edges, and 28,081 cell-type-to-receptor EXPRESSES
edges. Of the 1,541 receptor genes represented in the dataset, 1,538 were mapped
to existing HGNC gene nodes\cit{31}, producing approximately 129,234 new edges
in total.

These edges retain more information than a binary expression
relationship (Table~2).
Region-level and cell-type-level edges include the fraction of cells expressing
the receptor, expression magnitude, specificity, number of profiled cells,
receptor family, species, modality, and source dataset. The resulting profiles
reveal distinct patterns of receptor expression across human brain regions
(Figure~2B) and cell types (Figure~2C). Grouping receptors by neurotransmitter
system, including serotonergic, dopaminergic, glutamatergic, GABAergic,
adrenergic, cholinergic, opioid, and other neuromodulatory receptors, provides a
direct molecular interface to the circuit and cell-type relationships extracted
from the literature. Thus, once an agent identifies a disease-relevant region or
cell population, the graph can retrieve molecular effectors that are physically
available in the corresponding human biological context.

\begin{figure}[!htbp]
\centering
\widefig{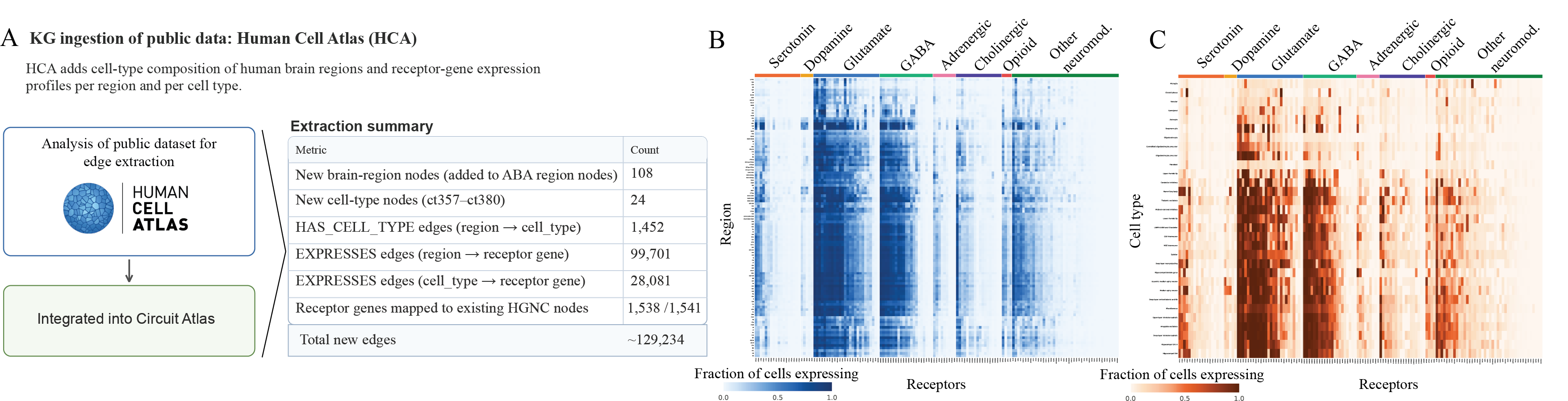}
\caption{\textbf{Integration of the Human Cell Atlas into CircuitATLAS.}
(A) Schematic of Human Cell Atlas data integration and summary of extracted
graph entities and relationships. The table shows the addition of 108
brain-region nodes, 24 cell-type nodes, 1,452 region-to-cell-type edges, 99,701
region-to-receptor expression edges, and 28,081 cell-type-to-receptor expression
edges. Of 1,541 receptor genes, 1,538 were mapped to existing HGNC nodes,
yielding approximately 129,234 new edges. (B, C) Heatmaps of receptor-gene
expression across (B) human brain regions and (C) cell types. Rows represent
brain regions or cell types, columns represent receptor genes, and color
intensity indicates the fraction of cells expressing each receptor. Receptor
genes are grouped by neurotransmitter system, as indicated above each heatmap.}
\label{fig:2}
\end{figure}

\begin{table}[!htbp]
\centering
\caption{\textbf{HCA-extract edges examples}}
\label{tab:2}
\scriptsize
\begin{tabularx}{\textwidth}{L{0.95cm} L{2.05cm} L{2.35cm} L{2.05cm} Y L{1.85cm} L{1.85cm}}
\toprule
\thd{Edge} & \thd{Head node} & \thd{Relation} & \thd{Tail node} &
\thd{Quantitative evidence} & \thd{Biological context} & \thd{Source}\\
\midrule
\textbf{HCA-1} & Human CA2U-CA3U, hippocampal CA2/3 & EXPRESSES &
GABRB3, GABA-A receptor $\beta$3 &
99.72\% of cells expressing; mean expression = 37.47; specificity = 0.081;
$n$ = 11,276 cells & Region-level aggregate; ligand-gated ion channel &
HCA Brain v1; Siletti et al., 2023\\
\addlinespace[2pt]
\textbf{HCA-2} & LAMP5-LHX6 and chandelier interneuron & EXPRESSES &
GRIN2A, NMDA receptor GluN2A &
99.8\% of cells expressing; mean expression = 42.98; specificity = 0.209;
$n$ = 500 cells & Human A13; ligand-gated ion channel &
HCA Brain v1; Siletti et al., 2023\\
\addlinespace[2pt]
\textbf{HCA-3} & Human cerebellum & HAS\ub CELL\ub TYPE & Cerebellar granule cell &
95.86\% of region cells assigned; $n$ = 23,783 cells &
Region-level cell-type composition & HCA Brain v1; Siletti et al., 2023\\
\bottomrule
\end{tabularx}
\end{table}

We next defined an equivalent ingestion structure for newly acquired \iv{} data
(Figure~3A). The current pipeline covers EEG, intracranial electrophysiology,
markerless pose estimation, fiber photometry, histology, and transcriptomics.
Each modality contributes different graph-native readouts. Electrophysiological
recordings generate cellular or circuit activity features, behavioral tracking
generates quantitative movement phenotypes, photometry connects neurotransmitter
or calcium dynamics to circuit features, histology links interventions and
disease models to cellular phenotypes, and transcriptomic analyses connect
experimental conditions to genes, transcripts, pathways, and cell types. Because
all modalities use the same canonical node vocabulary, observations collected in
different disease models and assays can be placed within a common evidence
space.

To illustrate this process, we selected three experimental datasets that
reproduce previously reported disease-associated phenotypes and converted their
findings into graph-native relationships. In the \textit{Fmr1} knockout model of
Fragile X syndrome, \iv{} electrophysiological recordings revealed increased
relative gamma-band power compared with wild-type controls (Figure~3B;
$P = 0.0098$, Hedges' $g = 1.17$), consistent with previous reports\cit{32}. In
the 6-OHDA model of Parkinson's disease, markerless behavioral tracking revealed
reduced open-field locomotor speed, providing a quantitative measure of motor
impairment (Figure~3C; $P = 0.0011$, Hedges' $g = -1.05$). In the acute 4-AP
seizure model, EEG recordings showed a within-animal increase in relative
gamma-band power following 4-AP administration compared with saline baseline
(Figure~3D; $P = 1\times10^{-5}$, Cohen's $d_z = 1.25$). Each observation was
represented as a relationship linking the corresponding disease model to an
electrophysiological or behavioral phenotype (Table~3), with associated metadata
specifying the direction of effect, statistical test, effect size, sample size,
assay, anatomical location, species, experimental condition, acquisition period,
and data provenance.

Together, these ingestion routes connect three complementary evidence layers:
published interpretations from the literature, human molecular anatomy from
public atlases, and newly generated functional measurements from experimental
models. The HCA data indicate which molecular effectors are expressed in
relevant human regions and cell types, while \iv{} data identify activity and
behavioral phenotypes that can be measured and modulated experimentally.
Representing both as provenance-linked graph evidence contextualizes new
findings against prior knowledge and creates cross-scale paths from
disease-associated phenotypes to circuits, cell populations and potentially
druggable molecular targets.

\begin{figure}[!htbp]
\centering
\widefig{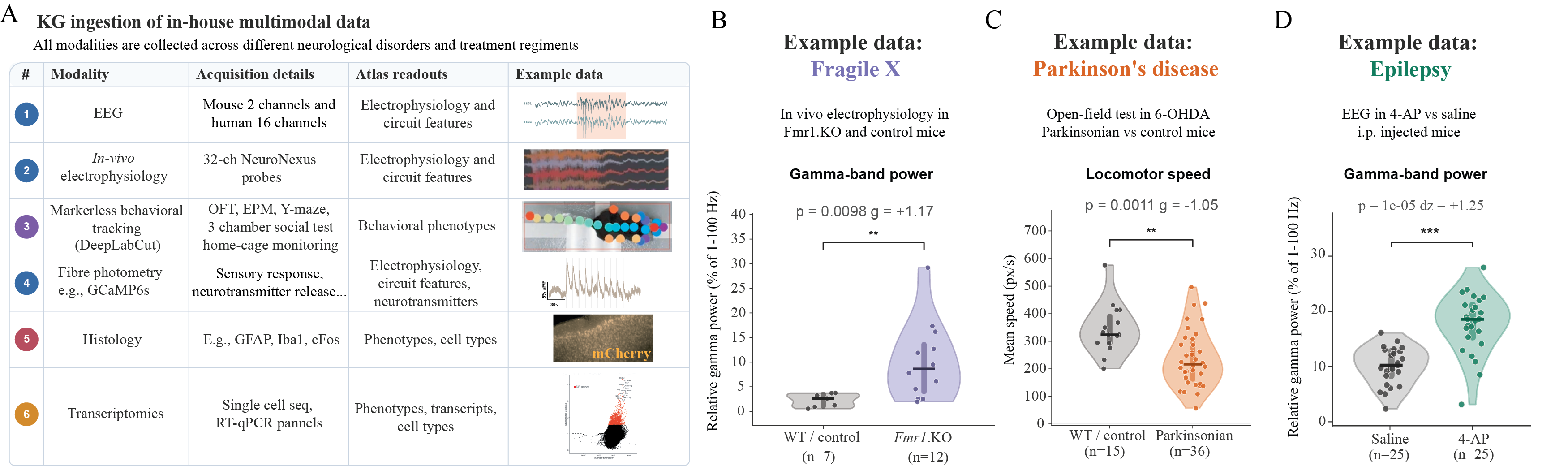}
\caption{\textbf{Integration of newly generated multimodal \textit{in vivo} data
into CircuitATLAS.}
(A) Overview of the six experimental modalities included in the in-house
data-ingestion pipeline, showing acquisition details, corresponding graph
readouts, and representative data for EEG, \textit{in vivo} electrophysiology,
markerless behavioral tracking, fiber photometry, histology, and
transcriptomics. (B) Relative gamma-band power in \textit{Fmr1} knockout mice
($n = 12$) and wild-type controls ($n = 7$; $P = 0.0098$, Hedges'
$g = 1.17$). (C) Mean open-field locomotor speed in 6-OHDA Parkinsonian mice
($n = 36$) and controls ($n = 15$; $P = 0.0011$, Hedges' $g = -1.05$).
(D) Relative gamma-band power following saline and 4-aminopyridine (4-AP)
administration in the same animals ($n = 25$; $P = 1\times10^{-5}$, Cohen's
$d_z = 1.25$).}
\label{fig:3}
\end{figure}

\begin{table}[!htbp]
\centering
\caption{\textbf{Representative edges extracted from in-house multimodal
\textit{in vivo} data}}
\label{tab:3}
\scriptsize
\begin{tabularx}{\textwidth}{L{0.85cm} L{1.95cm} L{2.3cm} L{2.05cm} Y L{2.2cm} L{2.2cm}}
\toprule
\thd{Edge} & \thd{Disease-model node} & \thd{Relation} &
\thd{Feature or phenotype node} & \thd{Experimental finding} &
\thd{Statistical evidence} & \thd{Assay and cohort}\\
\midrule
\textbf{E1} & 6-OHDA mouse (dm31) & RECAPITULATES &
Bradykinesia-related behavior (beh307) &
Open-field mean locomotor speed decreased from 324 to 216 px/s ($-33\%$) &
Two-sided Mann-Whitney U; $P$ = 0.0011; Hedges' $g$ = $-1.05$ &
DeepLabCut open-field tracking; $n$ = 15 controls and $n$ = 36 6-OHDA mice\\
\addlinespace[2pt]
\textbf{E2} & Fmr1 knockout mouse (dm249) & SHOWS &
Broadband gamma activity (cief54) &
Relative gamma power in S1 increased from 2.6\% to 8.7\% ($3.3\times$) &
Two-sided Mann-Whitney U; $P$ = 0.0098; Hedges' $g$ = $+1.17$ &
32-channel \iv{} LFP; $n$ = 7 wild-type and $n$ = 12 Fmr1 knockout mice\\
\addlinespace[2pt]
\textbf{E3} & 4-AP acute seizure model (dm617) & SHOWS &
Broadband gamma activity (cief54) &
Relative EEG gamma power increased from 10.2\% at saline baseline to 18.6\%
after 4-AP &
Paired Wilcoxon signed-rank; $P = 1.0\times10^{-5}$; Cohen's $d_z$ = $+1.25$ &
Overnight two-channel EEG; $n$ = 25 animals measured under both conditions\\
\bottomrule
\end{tabularx}
\end{table}

\clearpage

\section{A new agentic workflow identifies new therapies by reasoning from
circuit dysfunction to druggable effectors}

Circuit-based target discovery poses a different problem from conventional
molecular target identification. Rather than beginning with a gene or protein
altered in disease, it begins with a measurable pathological phenotype (e.g., an
oscillatory abnormality, altered excitability, or behavioral deficit) and asks
which intervention could move the underlying circuit in the desired direction.
This expands the therapeutic search space to molecular effectors that may be
entirely intact in the disease but provide useful control over the relevant cell
population or network. The same flexibility also makes the task difficult: a
plausible hypothesis must connect evidence across activity features, circuits,
regions, cell types, molecular mechanisms, therapeutic modalities, and clinical
constraints, while preserving the direction of each effect and avoiding
unsupported shortcuts between disease and target.

We therefore developed an agentic workflow modeled on the organization of a
translational systems neuroscience laboratory, in which specialized agents
coordinate across biological scales to generate and evaluate therapeutic
hypotheses (Figure~4). First, a coordinating PI agent converts an initial
disease-and-modality brief into a sequence of specialized searches. The circuit
agent identifies disease-relevant activity phenotypes and their anatomical and
cellular substrates; the molecular agent searches for effectors capable of
shifting those phenotypes while masking direct disease-molecule associations;
and modality and clinical agents evaluate whether the proposed mechanism can be
engaged feasibly and translated safely. Explicit gates require sourced,
directionally consistent evidence before a hypothesis can advance, while an
adversarial sceptic challenges each stage and can trigger renewed searches. This
decomposition is intended to retain the breadth of LLM-guided exploration while
making the resulting target hypotheses inspectable, falsifiable, and traceable
to graph evidence.

\begin{figure}[!htbp]
\centering
\includegraphics[height=0.70\textheight,keepaspectratio]{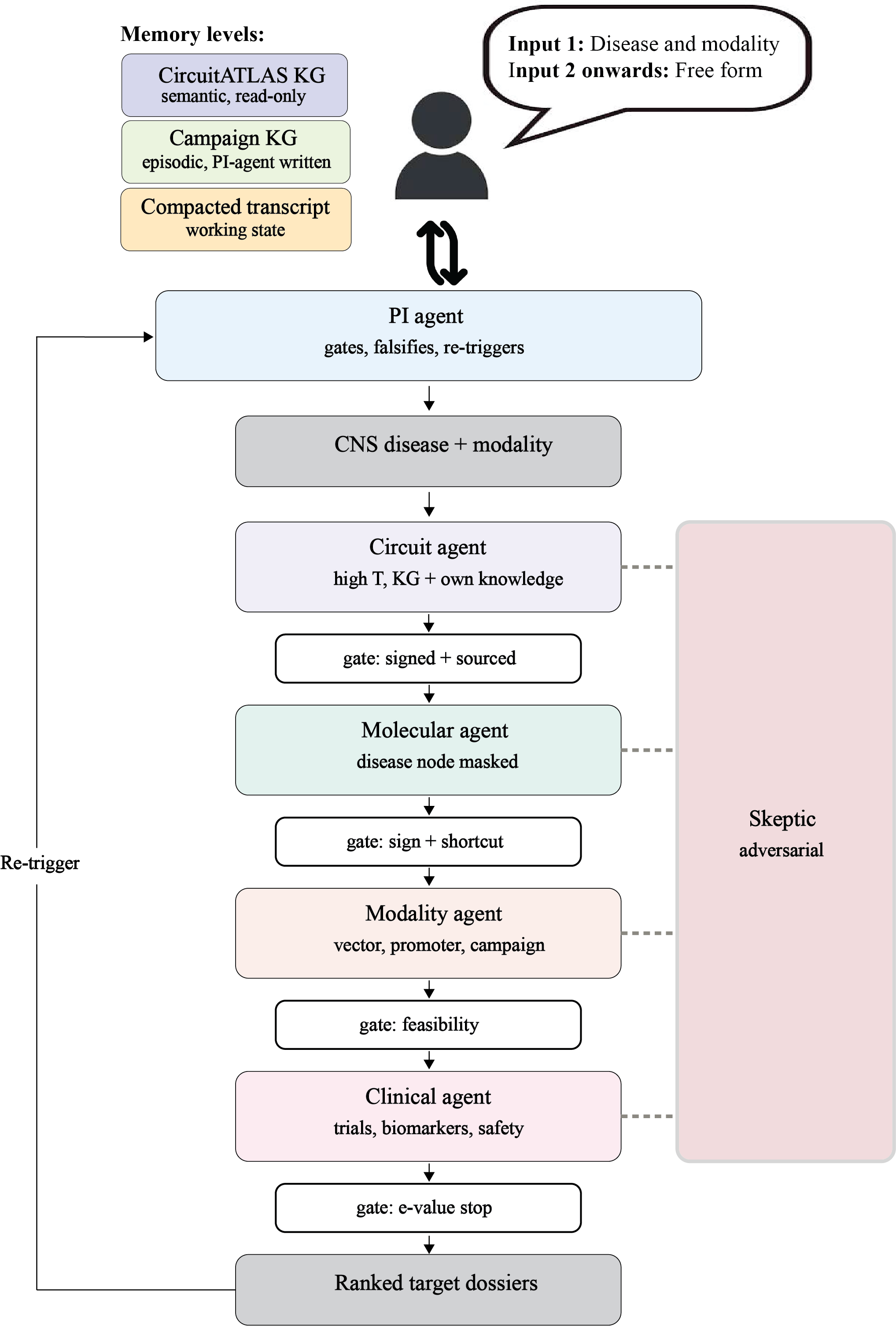}
\caption{\textbf{A circuit-first evidence-gated agentic workflow for cross-scale
target discovery.}
A user initiates a campaign with a central nervous system disease and
therapeutic modality, after which subsequent inputs may be provided in free
form. A coordinating PI agent manages the workflow, applies gates, identifies
failures, and re-triggers earlier stages when evidence is insufficient. The
circuit agent identifies disease-relevant activity phenotypes, circuits,
regions, and cell populations using CircuitATLAS. After a gate requiring sourced
and directionally coherent evidence, the molecular agent searches for effectors
capable of modulating the selected circuit phenotype while direct
disease-to-molecule information is masked to prevent shortcut reasoning. The
modality agent evaluates the feasibility of engaging the proposed target,
including intervention, vector, promoter, and campaign-level constraints, and
the clinical agent assesses existing trials, biomarkers, safety, and
translational evidence. An adversarial sceptic independently challenges the
reasoning produced at each specialist stage. Three memory layers maintain the
read-only semantic knowledge graph, an episodic campaign graph written by the PI
agent, and a compacted working transcript. The workflow terminates in ranked
target dossiers containing the supporting evidence, assumptions, uncertainties,
and proposed route to validation; dossiers that fail predefined evidence or
feasibility gates are rejected or returned for further search.}
\label{fig:4}
\end{figure}

To illustrate the hypotheses generated by this workflow, we applied the agent to
Fragile X syndrome, Dravet syndrome and epilepsy (Table~4). In Fragile X
syndrome, the workflow identified excessive thalamocortical gain and nominated
partial suppression of \textit{SLC17A6} (VGLUT2) in excitatory thalamic relay
neurons as a corrective intervention. Reducing vesicular glutamate loading was
predicted to lower quantal transmission at thalamocortical synapses and thereby
reduce cortical gain, although the essential role and broad subcortical
expression of VGLUT2 make partial and anatomically restricted modulation an
important constraint. In Dravet syndrome, the agent proposed partial suppression
of the somatodendritic \textit{HTR1A} autoreceptor in dorsal raphe serotonergic
neurons, with the aim of releasing these neurons from autoinhibition and
increasing forebrain serotonin tone. This strategy has clinical precedent in the
indication, since fenfluramine is approved for seizures associated with Dravet
syndrome and acts in part through serotonergic mechanisms, although the workflow
also surfaced potentially divergent effects of 5-HT1A signalling across seizure
phenotypes. In epilepsy, CircuitATLAS nominated \textit{ATP1A3}, the neuronal
$\alpha$3 Na$^{+}$/K$^{+}$-ATPase, as a control point on cortical excitability,
with potentiation predicted to increase activity-dependent hyperpolarizing
current and preserve inhibitory restraint during sustained firing. We selected
\textit{ATP1A3} for experimental validation and carried this hypothesis into the
\iv{} and structure-guided campaign described in the next section. Together,
these examples illustrate how CircuitATLAS identifies targets through their
capacity to control pathological circuit states rather than through their
molecular association with disease.

\begin{table}[!htbp]
\centering
\caption{\textbf{Example of three new targets identified by CircuitATLAS}}
\label{tab:4}
\scriptsize
\begin{tabularx}{\textwidth}{L{2.35cm} Y Y Y}
\toprule
& \thd{Hypothesis 1} & \thd{Hypothesis 2} & \thd{Hypothesis 3}\\
\midrule
\thd{Indication} & Fragile X syndrome & Dravet syndrome & Epilepsy\\
\addlinespace[2pt]
\thd{Circuit phenotype} & Thalamocortical hyperexcitability &
Lowered threshold for generalised synchronous discharge, from failed inhibitory
restraint &
Cortical hyperexcitability and failure to restrain sustained neuronal
firing\\
\addlinespace[2pt]
\thd{Measure / polarity} & Cortical evoked response amplitude and habituation
slope; resting gamma power; ASSR phase-locking. Higher amplitude / lower
habituation = greater gain\cit{33} &
Seizure and interictal discharge rate, hyperthermia seizure threshold. Higher
rate / lower threshold = greater excitability &
Seizure and interictal discharge rate; cortical activity during sustained
excitation. Higher activity / discharge rate = greater excitability\\
\addlinespace[2pt]
\thd{Corrective direction} & Decrease thalamocortical gain &
Decrease network excitability &
Decrease network excitability\\
\addlinespace[2pt]
\thd{Associated phenotypes} & Sensory hypersensitivity, impaired habituation,
audiogenic seizures, elevated resting gamma\cit{34}, reduced auditory
phase-locking\cit{35}, sleep spindle disruption, attention and arousal deficits,
anxiety &
Convulsive and myoclonic seizures, SUDEP via seizure-induced respiratory arrest,
developmental regression, ataxia &
Seizures, interictal discharges, impaired inhibitory restraint\\
\addlinespace[2pt]
\thd{Intervention logic} & Corrective: reduce the excess glutamatergic
thalamocortical drive &
Compensatory: recruit an intact neuromodulatory brake that is not itself
deficient &
Compensatory: augment an endogenous activity-dependent hyperpolarizing current
recruited during sustained firing\\
\addlinespace[2pt]
\thd{Region} & Thalamic relay neurons (excitatory) &
Dorsal raphe serotonergic neurons &
Cortical neurons\\
\addlinespace[2pt]
\thd{Molecular target} & \textit{SLC17A6} (VGLUT2) &
\textit{HTR1A} (somatodendritic 5-HT1A autoreceptor) &
\textit{ATP1A3} (neuronal $\alpha$3 Na$^{+}$/K$^{+}$-ATPase)\\
\addlinespace[2pt]
\thd{Direction on target} & Partial knockdown, $\sim$30--50\% &
Partial knockdown, $\sim$50--70\% &
Potentiation\\
\addlinespace[2pt]
\thd{Mechanism} & Lower vesicular glutamate loading \ra{} reduced quantal
amplitude at thalamocortical synapses \ra{} lower cortical gain &
Remove GIRK-mediated autoinhibition \ra{} raise tonic 5-HT firing \ra{} increase
forebrain 5-HT release \ra{} engage anticonvulsant 5-HT1A and 5-HT2A
signalling &
Increase $\alpha$3 pump capacity \ra{} greater Na$^{+}$ extrusion during
sustained firing \ra{} larger activity-dependent hyperpolarizing current \ra{}
reduce neuronal hyperexcitability\\
\addlinespace[2pt]
\thd{Modality} & AAV gene therapy with miRNA under VGlut2 promoter. Local
delivery or IV with BBB-crossing capsid plus validated promoter targeting. &
AAV gene therapy with miRNA under Tph2 promoter. Local delivery or IV with
BBB-crossing capsid plus validated promoter targeting. &
Small molecule. Structure-guided design against non-catalytic sites, with the
cardiotonic steroid pocket excluded as an anti-target.\\
\addlinespace[2pt]
\thd{Key risks} & VGLUT2 null is lethal, so degree titration is critical; broad
subcortical expression rules out systemic delivery &
Much DRN inhibitory input is extrinsic; 5-HT1A activation promotes spike-wave in
absence models, so mixed phenotypes may respond unevenly; anxiogenic potential &
Net circuit effect may be cell-type dependent, as increased pump activity could
also suppress firing in GABAergic interneurons; docking predicts binding, not
polarity; limited $\alpha$3 versus $\alpha$1 divergence may constrain
selectivity\\
\bottomrule
\end{tabularx}
\end{table}

\clearpage

\section{Closing the loop from target nomination to experimental validation}
\enlargethispage{\baselineskip}

A target-discovery system becomes useful when the hypotheses it generates can be
converted into experiments that resolve the uncertainties on which those
hypotheses depend. This is particularly difficult in \iv{} neuroscience, where
experiments are slow and hard to automatize, carry ethical considerations,
depend on finite specialist personnel, and must proceed under approved
protocols. We therefore represented our laboratory as a
machine-readable resource that an experiment-design agent can query when
planning the next step (Figure~5A). The Model Context Protocol server exposes 81
standard operating procedures together with operator competence, equipment
state, colony and consumable stock and reserved capacity as 25 tools and 20
resources. Proposed experiments are evaluated against statistical power,
expected information gain, 3Rs compliance and feasibility before human approval,
while those requiring unavailable capabilities are returned as capability
requests.

\begin{figure}[!htbp]
\centering
\includegraphics[height=0.68\textheight,keepaspectratio]{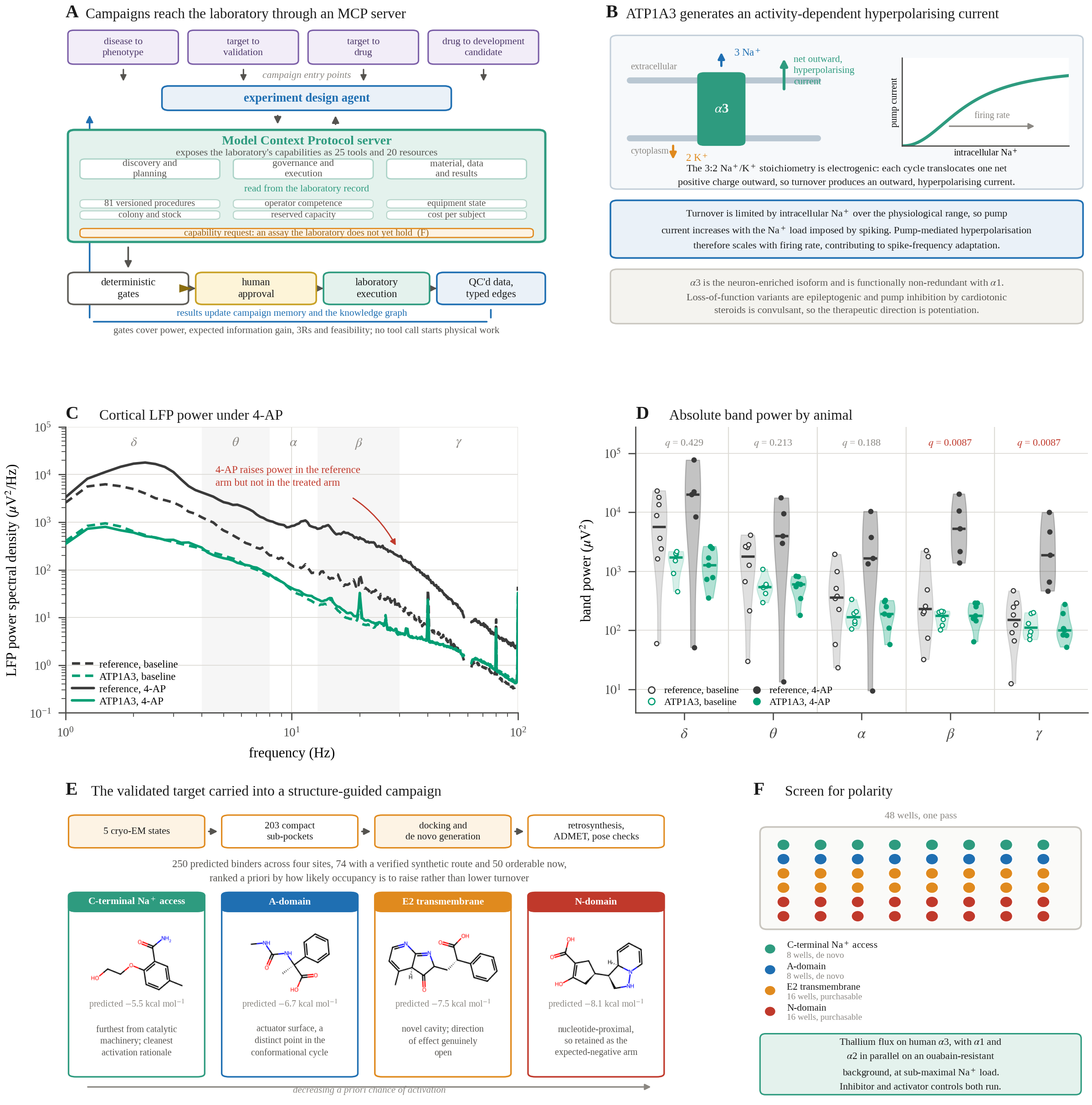}
\caption{\textbf{An \textit{in vivo} lab-in-the-loop, from a graph-derived
control point to a chemistry campaign.}
(A) Campaigns enter at one of four points and are planned by an
experiment-design agent that addresses the laboratory through a Model Context
Protocol server, exposing 25 tools and 20 resources over a structured record of
81 versioned procedures, operator competence, equipment state, colony and
consumable stock, reserved capacity and cost per subject. Designs pass
deterministic gates before human approval; an assay the laboratory does not
hold enters as a capability request.
(B) Pump turnover exports net positive charge and depends on intracellular
Na$^{+}$, so the hyperpolarizing current it generates is recruited in
proportion to the Na$^{+}$ that firing accumulates (inset).
(C) Mean cortical LFP power spectra for AAV9.mDlx.ATP1A3 and the reference
vector at baseline and under focal 4-AP, canonical bands shaded. Baseline power
differs between arms, so the primary contrast is within-animal.
(D) Absolute band power, one point per animal, with medians (reference baseline
$n = 8$, ATP1A3 baseline $n = 6$, reference 4-AP $n = 5$, ATP1A3 4-AP
$n = 7$); $q$ values are Benjamini--Hochberg corrected across bands for the
within-animal fold-change contrast.
(E) Structure-guided campaign on the validated target, with one representative
compound and predicted binding score per site, sites ordered by the prior
probability that occupancy raises turnover. Docking predicts binding, not
polarity.
(F) The polarity screen: 48 wells across the four site classes, thallium flux
on human $\alpha$3 with $\alpha$1 and $\alpha$2 in parallel on an
ouabain-resistant background, at sub-maximal Na$^{+}$ load.}
\label{fig:5}
\end{figure}

We used \textit{ATP1A3} to illustrate how a target hypothesis can be advanced
through this process. CircuitATLAS nominated the neuronal $\alpha$3
Na$^{+}$/K$^{+}$-ATPase as a control point on cortical excitability. Pump
turnover exports net positive charge and is recruited by the intracellular
Na$^{+}$ load that firing generates, producing an activity-dependent
hyperpolarizing current (Figure~5B)\cit{36}. ATP1A3 is restricted to neurons and
strongly expressed in GABAergic populations\cit{37}, while loss of function is
epileptogenic\cit{38} and pharmacological pump inhibition is
convulsant\cit{39}, together suggesting potentiation as the therapeutic
direction. The unresolved question was whether increasing ATP1A3 in inhibitory
interneurons would strengthen inhibition during sustained firing or instead
suppress their output.

We therefore expressed ATP1A3 in cortical GABAergic interneurons using
AAV9.mDlx.ATP1A3\cit{40} and recorded cortical local field potentials at
baseline and during focal 4-aminopyridine challenge (Figure~5C--D). No frequency
band differed significantly between arms at baseline. Under 4-AP, the median
within-animal fold change fell from 11.16 to 1.33 in beta,
9.73 to 1.19 in low gamma and 7.81 to 1.02 in high gamma, each at $q = 0.0087$
with a Cliff's $\delta$ of $-1.0$, while delta, theta and alpha were unchanged.
These results supported the hypothesis that increasing ATP1A3 in interneurons
blunts challenge-evoked cortical hyperexcitability, providing a basis for
advancing the target pharmacologically.

The next question was whether ATP1A3 could be modulated with a small molecule in
the required direction, so we carried the validated target into a
structure-guided campaign in Claude Science with Opus 5 (Figure~5E). The
campaign assembled five deposited human $\alpha$3 cryo-EM structures spanning
the Post-Albers cycle, re-derived 203 compact sub-pockets and combined docking,
structure-based generation, retrosynthetic route finding, ADMET prediction and
pose validation, excluding the inhibitory cardiotonic steroid site as an
anti-target. It returned 250 predicted binders across four sites, including 74
with a verified synthetic route and 50 directly orderable compounds, with sites
ranked by the prior probability that occupancy would increase rather than
decrease pump turnover. The modelling also exposed the principal uncertainties:
limited $\alpha$3 versus $\alpha$1 sequence divergence across most pockets, the
small size of the available cavities, and most importantly the
inability of docking alone to distinguish potentiators from inhibitors.

The campaign therefore converged on a functional polarity experiment as the next
step: compounds are tested by thallium flux on human $\alpha$3, with
$\alpha$1 and $\alpha$2 measured in parallel on an ouabain-resistant
background and at sub-maximal Na$^{+}$ load, followed for any hit by an ATPase
Na$^{+}$ dependence curve (Figure~5F). These assays were not in the laboratory
catalogue and were returned as capability requests. ATP1A3 thus
illustrates the broader experimental loop intended by CircuitATLAS, in which a
target hypothesis is advanced until its key uncertainty becomes experimentally
resolvable, the resulting evidence determines whether and how it should
progress, and each subsequent uncertainty defines the next informative
experiment.

\section{Discussion}

Here, we introduce CircuitATLAS, a framework for CNS target discovery that
begins with disease-associated neural activity phenotypes and identifies
molecular interventions capable of shifting pathological circuits toward more
functional states. We constructed a provenance-linked systems-neuroscience
knowledge graph integrating published literature, human molecular atlases, and
newly generated \iv{} experimental data. Building on this, we developed an
evidence-gated agentic workflow that connects neural activity phenotypes to
their circuit and cellular substrates and identifies molecular effectors capable
of modulating them. We further established a human-governed lab-in-the-loop
framework linking target nomination to experimental design, execution and
validation. Together, this establishes a therapeutic discovery approach in which
molecular targets are selected for their capacity to correct circuit
dysfunction, irrespective of whether they are themselves altered in disease.

Recent advances in protein structure prediction\cit{41}, molecular generation
and protein design\cit{42} are progressively lowering the barrier between a
specified biological target and a candidate therapeutic, making target
identification and validation an increasingly important bottleneck. As more of
the proteome becomes technically addressable, therapeutic discovery may become
less constrained by the availability of interventions than by our ability to
identify which biological nodes are most consequential to perturb. CircuitATLAS
provides a systematic way to prioritize these targets by integrating evidence
across biological scales.

This work also belongs to a broader shift from biomedical knowledge retrieval
toward scientific reasoning systems. Knowledge graphs such as SPOKE\cit{15},
TxGNN\cit{18} and more recent systems such as PROTON\cit{43} have demonstrated
the value of organizing heterogeneous biomedical relationships for prediction
and hypothesis generation. Agentic systems are increasingly extending this idea
from retrieval to multistep scientific workflows\cit{44,45}; ROBIN\cit{46}, for
example, coupled literature reasoning and experimental analysis to generate and
test biological hypotheses. CircuitATLAS differs principally in the
representation chosen for the problem. Electrophysiological features, circuits,
regions and cell populations are explicit intermediate objects rather than
contextual annotations around molecular disease associations. Our agentic
workflow then constrains reasoning through these intermediate levels, including
deliberately masking disease-molecule relationships during target generation to
reduce shortcut reasoning.

The potential of CircuitATLAS ultimately lies in combining literature-scale
knowledge (Figure~1) with large volumes of standardized experimental data
(Figures~2 and~3). Published studies provide a valuable record of existing scientific
knowledge, but are subject to publication bias, underreporting of negative
results, and incomplete documentation of the experimental conditions that
influence neural measurements. This is particularly consequential in systems
neuroscience, where minor sources of variation can change both neural and
pharmacological phenotypes\cit{47}. Addressing this variability requires more
standardized approaches to neural data generation, including consistent
acquisition hardware, experimental protocols, metadata collection, and analysis
pipelines, together with systematic reporting of negative results. Such
standardization would enable more direct comparisons of neural phenotypes and
intervention effects across disease models and experiments. CircuitATLAS
provides a common representation in which these newly generated observations can
be integrated with existing literature-derived evidence. As experimental
datasets accumulate, these initially heuristic weights could themselves be
calibrated against prospective outcomes, allowing the graph to learn not only
more biology, but which kinds of evidence are most predictive of successful
intervention.

Similarly, as agent-guided intervention-outcome datasets grow, the
\textit{lab-in-the-loop} could evolve into a genuine learning system.
Closed-loop laboratories in chemistry and protein engineering have already shown
how active learning or reinforcement-learning-style policies can use previous
experiments to select increasingly informative subsequent experiments\cit{48,49}.
In \iv{} neuroscience, however, noisy observations, slow experiments and ethical
costs make human oversight essential, which is why approval by a named
scientist sits between the agent's proposal and any animal work.

Several limitations remain. First, the central translational assumption is that
disease-relevant circuit phenotypes and their corrective directions are
sufficiently conserved between experimental models and humans. Human
systems-neuroscience datasets remain comparatively sparse and rarely permit
causal perturbation; molecular atlases\cit{30} can establish that candidate
effectors are present in the relevant human cells, but not that manipulating
them will produce the same circuit-level response across species. Second,
CircuitATLAS is constrained by the evidence from which it is built. The
literature is biased toward well-studied diseases, positive findings and
experimentally accessible mechanisms, while deterministic entity resolution and
relation extraction inevitably omit or misclassify some evidence. Although
provenance and inference-time verification make individual reasoning chains
auditable, they do not eliminate these limitations. Third, the performance of
the discovery system itself has not yet been established prospectively. The
examples presented here demonstrate that CircuitATLAS can generate biologically
plausible hypotheses and, for ATP1A3, nominate a target subsequently validated
\iv{}, but they do not establish its success rate, relative performance against
expert-led or alternative computational approaches, or ability to predict
therapeutic efficacy across indications. Several prospective, preregistered
campaigns will therefore be required to determine whether accumulating
experimental outcomes improve target prioritization and experimental efficiency
over time.

CircuitATLAS provides a framework for discovering therapeutic targets through
their capacity to control pathological neural circuits. By combining structured
knowledge, agentic reasoning and iterative experimentation, it offers a path
toward systematically identifying biological control points capable of restoring
brain function.

\section{Methods}

\subsection{CircuitATLAS ontology and entity vocabulary}

CircuitATLAS was designed as a heterogeneous graph spanning molecular, cellular,
circuit, organismal and translational neuroscience. The schema defines 23 node
classes, and 105 populated directed relation types (81 literature-derived, 24
canonical). Entity vocabularies were assembled from established resources where
possible: genes from NCBI\cit{23}, transcripts from Ensembl\cit{24}, proteins
from UniProt\cit{25}, signalling pathways from Reactome\cit{26}, brain regions
and neuronal cell types from the Allen Brain Atlas\cit{27}, and phenotypes from
the Human Phenotype Ontology\cit{28}. Concepts that are less consistently
represented in existing databases, including circuit motifs, brain states, and
cellular- and circuit-level electrophysiological features, were curated
manually. Each entity was assigned a canonical name, node class, synonym set and
unique CircuitATLAS identifier, while external identifiers were preserved when
available. Direct disease-to-gene and disease-to-protein relationships were
excluded from the target-discovery graph by design.

\subsection{Literature corpus and screening}

The literature corpus was collected from PubMed abstracts, PubMed Central and
Europe PMC open-access full text, and the bioRxiv, medRxiv and arXiv preprint
servers. We constructed a manually curated master neuroscience lexicon
containing 1,701 terms and applied a two-hit filter requiring at least two
lexicon terms within a document, retaining 15.9 million records. Retained documents were
then assigned deterministically to eight overlapping subcorpora based on
detected entities: disease, phenotype, behavior/cognition, brain region, cell
type, circuit motif, cellular electrophysiology and circuit electrophysiology.
Documents were identified by PMID, PMCID or DOI and deduplicated before
downstream processing.

\subsection{Candidate relation extraction and LLM verification}

Candidate relations were generated only between 81 predefined pairs of node
classes. Canonical names and synonyms were matched deterministically, and a
candidate evidence passage was generated when the corresponding head and tail
entities occurred within a 400-character window. Evidence was aggregated by
relation type and canonical node pair across documents, retaining up to three
supporting passages together with document and mention counts and publication
metadata. This stage generated 57.7 million candidate co-occurrences.

LLMs were subsequently used as constrained semantic verifiers rather than for
unrestricted relation generation. Each candidate was presented with the proposed
relation, head and tail entities and supporting passages, and verification was
executed as AWS Bedrock batch inference (10,000 records per job, up to 95
concurrent jobs). Amazon Nova 2 Lite served as the default verification tier,
and Claude Haiku 4.5 was applied as a second quality pass over the edges Nova 2
Lite had accepted. A subset of 4,050 edges, sampled as 50 edges per relation
type, was additionally re-judged with Claude Opus 5 using the same prompt given
to Nova 2 Lite and Haiku 4.5. The system message set three tasks: entity
linking; relation classification as supported, unsupported or uncertain,
where uncertain denotes that the head and tail entities co-occur but the
passage does not state the relationship, and is flagged as both common and a
correct verdict; and assignment of polarity and species. The prompt further
instructed the model to judge only what the passages say and not to use outside
knowledge to fill gaps, so that a claim the passage does not state is
uncertain however plausible it may be. Under this test, Opus 5 judged the
relationship not to be clearly stated in the supporting passages for 17.8\% of
the sampled edges. Models
returned a supported/unsupported/uncertain verdict together with an evidence span
and a set of structured attributes (polarity, directness, specificity, clarity,
methodology, species, sample size, effect size, causal evidence and manipulation
type). Accepted edges were resolved to canonical entities, filtered and
deduplicated: 26.8 million verified mention rows across 81 relation types
aggregated to 5,306,909 canonical literature edges. Each retained edge kept its
relation type, head and tail identifiers, verdict counts, supporting quotation,
source PMIDs, publication metadata and extraction provenance.

Canonical relationships, including gene-transcript-protein mappings, pathway
membership, protein interactions and drug-target associations, were imported
separately from structured databases and resolved onto the same entity
vocabulary, contributing 2,352,709 edges across 24 relation types. The resulting
graph contains 3,831,997 nodes and 7,659,618 edges; with the Human Cell Atlas
layer described below, the additional build contains 3,832,129 nodes and
7,788,852 edges. New documents can be processed independently and incorporated
through versioned incremental graph updates.

\subsection{Human Cell Atlas integration}

Adult human brain single-nucleus transcriptomic data (\cit{30}; neuronal and
non-neuronal h5ad matrices) were transformed into graph-native
region-cell-type and receptor-expression relationships. HCA brain regions and
cell types were resolved against existing CircuitATLAS entities, with unmatched
entities added to the vocabulary. Groups with fewer than 20 profiled cells were
discarded, and an edge was written when the receptor gene was detected in
$\geq 5\%$ of cells in that group. Each expression edge retained the fraction of
cells expressing the receptor, a specificity score defined as

\begin{equation}
\mathrm{specificity}
=\log_{2}\!\left(
\frac{\text{fraction in group}}{\text{fraction across all cells}}
\right),
\label{eq:specificity}
\end{equation}

the number of profiled cells, receptor family, species, modality, source dataset
and a three-level evidence-strength label (high: $\geq 25\%$ of cells and
$\geq 100$ cells; medium: $\geq 10\%$ and $\geq 50$; low: $\geq 5\%$ and
$\geq 20$). Absence of an edge therefore denotes a below-threshold combination
rather than a tested negative, and explicit negative edges can be generated by
joining the full gene $\times$ cell-type $\times$ region table against the
retained set. This pipeline added 108 brain-region nodes, 24 cell-type nodes,
1,452 HAS\_CELL\_TYPE edges, 99,701 region-to-receptor EXPRESSES edges and
28,081 cell-type-to-receptor EXPRESSES edges; 1,538 of 1,541 receptor genes
mapped to existing HGNC gene nodes\cit{31} and no new gene entities were
created.

\subsection{Integration of \textit{in vivo} experimental data}

Experimental analysis outputs were converted into the same typed edge
representation used for literature and atlas evidence. Supported modalities
included EEG, intracranial electrophysiology, markerless behavioral tracking,
fiber photometry, histology and transcriptomics. Each observation was written as
a canonical head-relation-tail triple for example disease\_model \ra{} behavior
(RECAPITULATES) and disease\_model \ra{} circuit\_electrophysiology\_feature
(SHOWS); and retained its direction of effect, statistical test, $P$-value,
effect size, sample size, assay, brain region, species, strain or genotype,
acquisition dates and data provenance.

Intracranial recordings were acquired on 32-channel NeuroNexus probes\cit{50}
(Open Ephys, 30 kHz). Line noise was auto-detected per recording by
50-versus-60 Hz peak prominence and notched with three harmonics; the spike band
(300-6,000 Hz) was common-median referenced and the LFP band (1-300 Hz)
resampled to 1 kHz. Bad channels were detected per recording and excluded, the
first 5 s of each recording was discarded, and band power was computed over
$\delta$ 1-4, $\theta$ 4-8, $\alpha$ 8-13, $\beta$ 13-30, $\gamma$ 30-100 Hz.
EEG was analysed by Welch PSD\cit{51} (2-s Hann windows, 50\% overlap) after
skipping the first 45 min, with mains frequencies interpolated out and gamma
expressed as a percentage of 1--100 Hz total power. Behavior was tracked with
DeepLabCut\cit{52} (30 fps).

Representative analyses were: open-field locomotor speed in 6-OHDA versus
control mice (median $324 \rightarrow 216$ px/s; Mann--Whitney $U = 428$,
$P = 0.0011$, Hedges' $g = -1.05$; $n = 15$ versus 36); relative gamma power in
Fmr1 knockout versus wild-type S1 ($2.6\% \rightarrow 8.7\%$; $U = 12$,
$P = 0.0098$, $g = +1.17$; $n = 7$ versus 12); and paired EEG relative gamma
power at baseline and after 4-AP administration
($10.2\% \rightarrow 18.6\%$; Wilcoxon signed-rank $W = 16$,
$P = 1.0\times10^{-5}$, Cohen's $d_z = +1.25$; $n = 25$ paired animals).
Analyses were performed at the animal level, using the earliest available
session per animal where a gene-therapy arm would otherwise confound the
comparison.

\subsection{Agentic workflow}

Our agentic reasoning workflow was implemented as a staged, evidence-gated
reasoning workflow running Claude Opus 5 on Amazon Bedrock, with stage-specific
sampling temperatures (0.2-0.9). A coordinating PI agent decomposed an initial
disease and therapeutic-modality brief into specialist stages. The circuit agent
proposed disease-associated activity features, anatomical regions, circuits and
cell populations; substrates advanced only if their measure was directly
assayable, carried an explicit polarity statement, and had an
excitatory/inhibitory class consistent with the stated phenotype direction. The
molecular stage searched for effectors capable of moving the selected circuit
phenotype, retrieving through two channels: channel A, restricted to targets
with atlas-supported expression in the substrate, and channel B, proposing
effectors to be introduced ectopically. Modality and clinical agents
subsequently evaluated intervention feasibility, targeting strategy, clinical
precedent, known class hazards and off-target populations. An adversarial
sceptic evaluated every intermediate hypothesis.

Three design invariants were enforced in code rather than by prompt. First,
discovery is phenotype-down: gene-disease association (direct or indirect, e.g.,
through disease-model) never constitutes a reward signal. Second, the molecular
stage is disease-blind by construction, it queries a masked view in which the
disease class, its resolved disease models, eponym-named nodes, class-specific
mediators and low-disease-degree proxy mediators are removed at the query layer,
and in which mediator node classes are de-identified to an opaque
type-plus-identifier reference while molecular targets and the circuit substrate
retain their names. Third, sign is first-class: the required intervention
direction is computed as the product of corrective direction, excitability sign
and circuit sign, and inconsistencies are rejected deterministically.

The mask's acceptance test was adversarial re-identification: the exact masked
payload was handed to a model with no other context and asked to name the
disease. Under this test the specific disease was not recovered, but in some
cases a stronger multi-agent attack did recover the disease class (e.g.,
recovered epilepsy but not Dravet syndrome) from the retained circuit substrate.
Because the circuit phenotype is class-defining and removing it would defeat
phenotype-down discovery, we accept this and state the claim precisely: not that
the agent is always blind to the disease class, but that every candidate's
evidence chain is disease-free by construction and reconstructible from the
masked subgraph alone. Class-level identifiability was recorded as a campaign
property when it occurred.

Candidates passed through an independent auditor of deterministic gates covering
provenance grading, shortcut exclusion and Class-A novelty (dual Open Targets
and Europe PMC signals, reported separately), sign composition, level-jump
completeness, sign consistency, structural viability and population scope, with
an anytime-valid e-value stopping rule. Gates enforced: LLM judgement only
scores, the sceptic assigns severity and never holds a veto. Hypotheses were
assigned to one of five mechanism templates (cell-autonomous excitability;
transmitter supply; ectopic receptor; extrinsic/glial/metabolic;
synaptic/structural) and a deterministic payload tier (direct ionic conductance
$>$ transporter/enzyme $>$ GPCR). Outputs were labelled SUPPORTED (every chain
step evidence-anchored), PLAUSIBLE (coherent chain with gaps named) or EXCLUDED
(structurally impossible), with only EXCLUDED hypotheses withheld from the
report.

Three memory layers were maintained: the read-only CircuitATLAS semantic graph
accessed through the masked view; an append-only, campaign-specific episodic
store in which superseded entries are linked rather than overwritten and
rejections persist as forbidden repeats; and a compacted working transcript.
Final outputs were structured target dossiers containing the selected phenotype
and its polarity, corrective direction, circuit and cellular substrate,
molecular target and mechanistic chain, intervention modality and its
trade-offs, novelty at gene and mechanism level, named evidence gaps, a
pre-mortem failure analysis and a falsifiable validation package.

In a three-disease campaign (Fragile X syndrome, Dravet syndrome and epilepsy),
our agent reported 29 dossiers per disease, collapsing to 16, 17 and 15 distinct
hypotheses respectively. Of which five hypotheses were EXCLUDED on structural
grounds, including sodium-channel blockade in a seizure disorder.

\subsection{Laboratory interface and experiment governance}

The laboratory is exposed to the experiment-design agent through a Model Context
Protocol server (Figure~5A). The server reads a structured laboratory record
holding 81 versioned procedures together with their controlled documents,
operator competence per procedure, equipment state, animal colony and consumable
stock, reserved capacity and measured cost per subject, and projects it as 25
tools and 20 resources covering retrieval of laboratory state, validation of a
proposed design against each procedure's input schema, scheduling, and the
experiment lifecycle. The agent therefore plans against the procedures the
laboratory can currently execute rather than against the space of procedures
described in the literature, and a campaign may enter at disease to phenotype,
target to validation, target to drug, or drug to development candidate.

Proposed designs are evaluated by deterministic gates for statistical power,
expected information gain, 3Rs compliance and feasibility before being
submitted. A design that fails a gate is returned rather than queued.
Submission places an experiment in a review queue and does not initiate physical
work; approval, scheduling, quality-control adjudication and data release each
require an explicit decision recorded against a named operator, after which the
resulting data are returned to the agent. An assay the laboratory does not hold
is returned through the same interface as a capability request rather than
substituted.

\subsection{Interneuron-restricted ATP1A3 \textit{in vivo} validation}

ATP1A3 was packaged as AAV9.mDlx.ATP1A3 to restrict expression predominantly to
GABAergic interneurons and delivered to mouse somatosensory barrel cortex, with
AAV9.mDlx.mCherry delivered by the same route as the reference vector. Cortical
local field potentials were recorded \iv{} in awake head-fixed animals using
32-channel NeuroNexus silicon probes\cit{50}, at baseline and under acute focal
4-aminopyridine challenge, and were preprocessed as described above. Power was
summarised per animal in the delta, theta, alpha, beta, low gamma and high gamma
bands.

Absolute baseline power differed between arms, so the primary contrast was the
within-animal fold change in band power under 4-AP relative to that animal's own
baseline. Arms were compared by
two-sided Mann--Whitney test with Benjamini--Hochberg correction across the six
bands, and effect size was reported as Cliff's $\delta$. Baseline power was
compared between arms using the same test and correction. The fold-change
contrast used the animals contributing both conditions, five in the reference
arm and six in the ATP1A3 arm. Figure~5D displays absolute band power for every
recording in each condition (reference baseline $n = 8$, ATP1A3 baseline
$n = 6$, reference 4-AP $n = 5$, ATP1A3 4-AP $n = 7$).

\subsection{Structure-guided small-molecule campaign}

The chemistry campaign was run in Claude Science with Opus 5 (Figure~5E).
Five deposited human $\alpha$3 Na$^{+}$/K$^{+}$-ATPase cryo-EM structures
spanning the Post-Albers cycle were assembled, pockets were detected and
re-derived into 203 compact sub-pockets, and candidate ligands were produced and
filtered by docking, structure-based generation, retrosynthetic route finding,
ADMET prediction and pose-geometry validation. The inhibitory cardiotonic
steroid pocket was defined as an anti-target and excluded throughout.
Sub-pockets were annotated for isoform selectivity by comparing $\alpha$3 and
$\alpha$1 residues at lining positions, and for synthetic tractability by
heavy-atom count; 175 of the 203 contain no non-conservative substitution
between the two isoforms, and cavity volumes of 98 to 130~\AA$^{3}$ admit only
fragment-scale ligands, for which retrosynthetic success falls from 76\% at
twelve or fewer heavy atoms to 6\% at seventeen to twenty. The campaign returned
250 predicted binders across four sites, 74 with a verified synthetic route and
50 orderable without synthesis, and the sites were ranked a priori by the
likelihood that occupancy raises rather than lowers turnover. The tip-to-tip
separation between D887 and R901 on the extracellular DR loop was measured
across all five catalytic states.

Because docking returns a predicted binding score and does not distinguish a
potentiator from an inhibitor at the same site, the campaign terminates in a
polarity assay rather than a lead (Figure~5F): 48 wells distributed across the
four site classes, read by thallium flux on human $\alpha$3 with $\alpha$1 and
$\alpha$2 run in parallel on an ouabain-resistant background and at sub-maximal
Na$^{+}$ load, with inhibitor and activator controls. Compounds modulating in
the intended direction proceed to an ATPase Na$^{+}$ dependence curve, which
separates a left shift in apparent Na$^{+}$ affinity from a uniform increase in
turnover. Neither assay is in the current laboratory catalogue and both were
submitted as capability requests through the interface described above.

\section*{Ethics statement}

Experiments were approved by the local ethical review committee at Exin
Therapeutics and performed in accordance with local regulations. Mice of either
sex were generated via targeted breeding and housed in a temperature
(20--24\textdegree{} Celsius) and humidity (40\%)-controlled room under a 12 h
dark/light cycle with free access to food and water \textit{ad libitum}.

\section*{Data availability}

The code and the data supporting the findings of this study will be made
publicly available upon peer-reviewed publication.

\section*{Author contributions}

Gabriel Ocana-Santero conceived the study, designed and built the knowledge
graph and the agentic workflows, performed the analyses, prepared the figures
and wrote the manuscript. Marko Tvrdic contributed to the
conception of the study and to reviewing and editing the manuscript. Large
language models were used to assist with the writing and editing of this
manuscript; the authors reviewed and verified all content and take full
responsibility for it.

\section*{Competing interests}

The authors declare the following competing interests. Gabriel Ocana-Santero is
a co-founder, Chief Executive Officer and shareholder of Exin Therapeutics,
Inc., the company at which this work was carried out. Marko Tvrdic is a
co-founder, Chief Operating Officer and shareholder of Exin Therapeutics, Inc.
A provisional patent application relating to the work described here has been
filed by Exin Therapeutics, Inc.

\section*{References}
\addcontentsline{toc}{section}{References}

\begingroup
\small
\begin{enumerate}[label=\arabic*., leftmargin=2.2em, labelsep=0.5em,
                  itemsep=1pt, topsep=4pt, parsep=0pt]

\item Deisseroth, K. Circuit dynamics of adaptive and maladaptive behaviour.
\textit{Nature} 505, 309--317 (2014).
\item Bassett, D. S. \& Sporns, O. Network neuroscience. \textit{Nat. Neurosci.}
20, 353--364 (2017).
\item Ocana-Santero, G. \textit{et al.} Perinatal serotonin signalling
dynamically influences the development of cortical GABAergic circuits with
consequences for lifelong sensory encoding. \textit{Nat. Commun.} 16, 5203
(2025).
\item Uhlhaas, P. J. \& Singer, W. Neural Synchrony in Brain Disorders:
Relevance for Cognitive Dysfunctions and Pathophysiology. \textit{Neuron} 52,
155--168 (2006).
\item Paz, J. T. \& Huguenard, J. R. Microcircuits and their interactions in
epilepsy: Is the focus out of focus? \textit{Nat. Neurosci.} 18, 351--359
(2015).
\item Ligaard, J., Sann\ae s, J. \& Pihlstr\o m, L. Deep brain stimulation and
genetic variability in Parkinson's disease: a review of the literature.
\textit{NPJ Park. Dis.} 5, 18 (2019).
\item Santos, R. \textit{et al.} A comprehensive map of molecular drug targets.
\textit{Nat. Rev. Drug Discov.} 16, 19--34 (2017).
\item Ocana-Santero, G., D\'iaz-Nido, J. \& Herranz-Mart\'in, S. Future
prospects of gene therapy for Friedreich's ataxia. \textit{Int. J. Mol. Sci.}
22, 1815 (2021).
\item Gribkoff, V. K. \& Kaczmarek, L. K. The Need for New Approaches in CNS
Drug Discovery: Why Drugs Have Failed, and What Can Be Done to Improve Outcomes.
\textit{Neuropharmacology} 120, 11--19 (2017).
\item L\"oscher, W. \& Klein, P. The Pharmacology and Clinical Efficacy of
Antiseizure Medications: From Bromide Salts to Cenobamate and Beyond.
\textit{CNS Drugs} 35, 935--963 (2021).
\item Deuschl, G. \textit{et al.} A Randomized Trial of Deep-Brain Stimulation
for Parkinson's Disease. \textit{N. Engl. J. Med.} 355, 896--908 (2006).
\item Morrell, M. J. \& RNS System in Epilepsy Study Group. Responsive cortical
stimulation for the treatment of medically intractable partial epilepsy.
\textit{Neurology} 77, 1295--1304 (2011).
\item Hunt, R. F., Girskis, K. M., Rubenstein, J. L., Alvarez--Buylla, A. \&
Baraban, S. C. GABA progenitors grafted into the adult epileptic brain control
seizures and abnormal behavior. \textit{Nat. Neurosci.} 16, 692--697 (2013).
\item Wykes, R. C. \textit{et al.} Optogenetic and potassium channel gene
therapy in a rodent model of focal neocortical epilepsy. \textit{Sci. Transl.
Med.} 4, 161ra152 (2012).
\item Morris, J. H. \textit{et al.} The scalable precision medicine open
knowledge engine (SPOKE): a massive knowledge graph of biomedical information.
\textit{Bioinformatics} 39, btad080 (2023).
\item Chandak, P., Huang, K. \& Zitnik, M. Building a knowledge graph to enable
precision medicine. \textit{Sci. Data} 10, 67 (2023).
\item Himmelstein, D. S. \textit{et al.} Systematic integration of biomedical
knowledge prioritizes drugs for repurposing. \textit{eLife} 6, e26726 (2017).
\item Huang, K. \textit{et al.} A foundation model for clinician-centered drug
repurposing. Preprint at \url{https://doi.org/10.1101/2023.03.19.23287458}
(2024).
\item Kilicoglu, H., Shin, D., Fiszman, M., Rosemblat, G. \& Rindflesch, T. C.
SemMedDB: a PubMed-scale repository of biomedical semantic predications.
\textit{Bioinformatics} 28, 3158--3160 (2012).
\item Wei, C.-H. \textit{et al.} PubTator 3.0: an AI-powered literature resource
for unlocking biomedical knowledge. \textit{Nucleic Acids Res.} 52, W540--W546
(2024).
\item Lewis, P. \textit{et al.} Retrieval-Augmented Generation for
Knowledge-Intensive NLP Tasks. Preprint at
\url{https://doi.org/10.48550/arXiv.2005.11401} (2021).
\item Skarlinski, M. D. \textit{et al.} Language agents achieve superhuman
synthesis of scientific knowledge. Preprint at
\url{https://doi.org/10.48550/arXiv.2409.13740} (2024).
\item Sayers, E. W. \textit{et al.} Database resources of the national center
for biotechnology information. \textit{Nucleic Acids Res.} 50, D20--D26 (2022).
\item Dyer, S. C. \textit{et al.} Ensembl 2025. \textit{Nucleic Acids Res.} 53,
D948--D957 (2025).
\item The UniProt Consortium. UniProt: the Universal Protein Knowledgebase in
2025. \textit{Nucleic Acids Res.} 53, D609--D617 (2025).
\item Milacic, M. \textit{et al.} The Reactome Pathway Knowledgebase 2024.
\textit{Nucleic Acids Res.} 52, D672--D678 (2024).
\item Lein, E. S. \textit{et al.} Genome-wide atlas of gene expression in the
adult mouse brain. \textit{Nature} 445, 168--176 (2007).
\item K\"ohler, S. \textit{et al.} The Human Phenotype Ontology in 2021.
\textit{Nucleic Acids Res.} 49, D1207--D1217 (2021).
\item Geirhos, R. \textit{et al.} Shortcut learning in deep neural networks.
\textit{Nat. Mach. Intell.} 2, 665--673 (2020).
\item Siletti, K. \textit{et al.} Transcriptomic diversity of cell types across
the adult human brain. \textit{Science} 382, eadd7046 (2023).
\item Seal, R. L. \textit{et al.} Genenames.org: the HGNC resources in 2023.
\textit{Nucleic Acids Res.} 51, D1003--D1009 (2023).
\item Lovelace, J. W., Ethell, I. M., Binder, D. K. \& Razak, K. A.
Translation-relevant EEG phenotypes in a mouse model of Fragile X Syndrome.
\textit{Neurobiol. Dis.} 115, 39--48 (2018).
\item Ethridge, L. E. \textit{et al.} Reduced habituation of auditory evoked
potentials indicate cortical hyper-excitability in Fragile X Syndrome.
\textit{Transl. Psychiatry} 6, e787 (2016).
\item Wang, J. \textit{et al.} A resting EEG study of neocortical
hyperexcitability and altered functional connectivity in fragile X syndrome.
\textit{J. Neurodev. Disord.} 9, 11 (2017).
\item Ethridge, L. E. \textit{et al.} Neural synchronization deficits linked to
cortical hyper-excitability and auditory hypersensitivity in fragile X syndrome.
\textit{Mol. Autism} 8, 22 (2017).
\item Kim, J. H., Sizov, I., Dobretsov, M. \& von Gersdorff, H. Presynaptic
Ca$^{2+}$ buffers control the strength of a fast post-tetanic hyperpolarization
mediated by the $\alpha$3 Na$^{+}$/K$^{+}$-ATPase. \textit{Nat. Neurosci.} 10,
196--205 (2007).
\item B\o{}ttger, P. \textit{et al.} Distribution of Na/K-ATPase alpha 3 isoform,
a sodium-potassium P-type pump associated with rapid-onset of dystonia
parkinsonism (RDP) in the adult mouse brain. \textit{J. Comp. Neurol.} 519,
376--404 (2011).
\item Clapcote, S. J. \textit{et al.} Mutation I810N in the $\alpha$3 isoform of
Na$^{+}$,K$^{+}$-ATPase causes impairments in the sodium pump and
hyperexcitability in the CNS. \textit{Proc. Natl. Acad. Sci. U.S.A.} 106,
14085--14090 (2009).
\item Vaillend, C., Mason, S. E., Cuttle, M. F. \& Alger, B. E. Mechanisms of
neuronal hyperexcitability caused by partial inhibition of
Na$^{+}$-K$^{+}$-ATPases in the rat CA1 hippocampal region.
\textit{J. Neurophysiol.} 88, 2963--2978 (2002).
\item Dimidschstein, J. \textit{et al.} A viral strategy for targeting and
manipulating interneurons across vertebrate species. \textit{Nat. Neurosci.} 19,
1743--1749 (2016).
\item Jumper, J. \textit{et al.} Highly accurate protein structure prediction
with AlphaFold. \textit{Nature} 596, 583--589 (2021).
\item Watson, J. L. \textit{et al.} De novo design of protein structure and
function with RFdiffusion. \textit{Nature} 620, 1089--1100 (2023).
\item Noori, A. \textit{et al.} Graph AI generates neurological hypotheses
validated in molecular, organoid, and clinical systems. Preprint at
\url{https://doi.org/10.48550/arXiv.2512.13724} (2025).
\item Boiko, D. A., MacKnight, R., Kline, B. \& Gomes, G. Autonomous chemical
research with large language models. \textit{Nature} 624, 570--578 (2023).
\item Ifargan, T., Hafner, L., Kern, M., Alcalay, O. \& Kishony, R. Autonomous
LLM-driven research from data to human-verifiable research papers. Preprint at
\url{https://doi.org/10.48550/arXiv.2404.17605} (2024).
\item Ghareeb, A. E. \textit{et al.} A multi-agent system for automating
scientific discovery. \textit{Nature} 655, 497--505 (2026).
\item Georgiou, P. \textit{et al.} Experimenters' sex modulates mouse behaviors
and neural responses to ketamine via corticotropin releasing factor.
\textit{Nat. Neurosci.} 25, 1191--1200 (2022).
\item Burger, B. \textit{et al.} A mobile robotic chemist. \textit{Nature} 583,
237--241 (2020).
\item MacLeod, B. P. \textit{et al.} Self-driving laboratory for accelerated
discovery of thin-film materials. \textit{Sci. Adv.} 6, eaaz8867 (2020).
\item Siegle, J. H. \textit{et al.} Open Ephys: an open-source, plugin-based
platform for multichannel electrophysiology. \textit{J. Neural Eng.} 14, 045003
(2017).
\item Welch, P. The use of fast Fourier transform for the estimation of power
spectra: A method based on time averaging over short, modified periodograms.
\textit{IEEE Trans. Audio Electroacoustics} 15, 70--73 (1967).
\item Mathis, A. \textit{et al.} DeepLabCut: markerless pose estimation of
user-defined body parts with deep learning. \textit{Nat. Neurosci.} 21,
1281--1289 (2018).

\end{enumerate}
\endgroup

\end{document}